\documentstyle[preprint,prd,floats,aps]{revtex}
\newcommand{\be}{\begin{equation}}
\newcommand{\ee}{\end{equation}}
\newcommand{\ba}{\begin{eqnarray}}
\newcommand{\ea}{\end{eqnarray}}
\newcommand{\nr}[1]{(\ref{#1})}
\newcommand{\la}[1]{\label{#1}}

\newcommand{\msbar}{\overline{\mbox{\rm MS}}}
\newcommand{\vali}{\vspace*{0.5cm}}
\def\lsim{\raise0.3ex\hbox{$<$\kern-0.75em\raise-1.1ex\hbox{$\sim$}}}
\def\gsim{\raise0.3ex\hbox{$>$\kern-0.75em\raise-1.1ex\hbox{$\sim$}}}
 
\begin{document}
\setlength{\baselineskip}{0.6cm}
\draft
\title{
\mbox{ } \\
\mbox{ } \\
Comparison of 4d and 3d Lattice Results for \\
the Electroweak Phase Transition}
\author{
M. Laine\thanks{m.laine@thphys.uni-heidelberg.de}}
\address{
Institut f\"ur Theoretische Physik, 
Philosophenweg 16, 
D-69120 Heidelberg, Germany}
%\date{\today}
\maketitle
\vspace*{-0.8cm}
 
\begin{abstract}
\setlength{\baselineskip}{0.6cm}
We compare 4d lattice results for the finite temperature phase
transition in the SU(2)+Higgs model with 3d lattice results for the
phase transition in the corresponding dimensionally reduced effective
theory.  While the large errorbars and the lack of a relation of the 4d 
lattice gauge coupling to continuum physics prevent rigorous conclusions, 
the results are nevertheless compatible. This provides a direct
non-perturbative check of dimensional reduction in the present context.
\end{abstract}
\narrowtext
 
\vspace*{-12.6cm}
\noindent
\hspace*{13.0cm} \mbox{HD-THEP-96-08} \\
\hspace*{13.0cm} \mbox{hep-lat/9604011} \\
\hspace*{13.0cm} April 12, 1996 %\today
\vspace*{9.5cm}
 
\section*{}
 
{\bf 1.} Due to its effect on the baryon number of the Universe, 
the electroweak phase transition in the Standard Model 
and beyond is of considerable interest
(for a review, see~\cite{rs}). 
The first question to be asked is whether the phase
transition is strongly enough of first order for cosmological
consequences. %~\cite{krs}. 
It turns out that even this simple question 
is difficult to answer reliably, due to the infrared (IR) problem
in finite temperature field theory. %~\cite{ir}. 
This has led to several lattice investigations of the 
problem~[2--12].
  
There are two different lattice approaches available. 
First, there are the finite temperature 4d simulations of
the bosonic SU(2)+Higgs theory~[2--6]. %\cite{bunk,des1,des3,des4}.
Second, there are 3d 
simulations~[7--12] %\cite{fkrs,klrs2,leip1,leip2,ptw,knpr}
of an effective theory obtained with dimensional 
reduction~\cite{jkp,fkrs1} from the original theory. 
Recently, the extrapolation to the continuum limit
has been investigated both in 4d~\cite{des4} and in 3d~\cite{klrs2}. 
Hence, a comparison of the results should become possible,
allowing to check non-perturbatively the validity of 
the perturbative dimensional reduction used in the derivation 
of the effective 3d theory. 
We will make the comparison in a part of the parameter space
where corrections to the 3d theory are largest
and where the transition is strong enough to be accessible
also to the 4d simulations. However, the corrections are still very
small and hence the 4d and 3d results turn out to 
agree within errorbars. 
\vali

{\bf 2.} Let us start by briefly reviewing the main features
of the two approaches. In principle, the more straightforward
way of solving the problem is the 4d approach, in which the 
original theory is put as such on the lattice. 
The parameters of the lattice action are fixed from masses
and from a suitably defined coupling constant measured at zero temperature. 
Then the lattice shape is changed so as to go finite
temperature. 

Unfortunately, there are a few problems
with the 4d approach. First, one cannot put the chiral fermions
appearing e.g. in the Standard Model on lattice, and hence
one can only study the bosonic sector of the theory. This is 
unacceptable, since especially the heavy top quark is numerically 
important for the electroweak phase transition. Second, 
it is not known how the
renormalized lattice gauge coupling $g_R^2$ 
of the 4d simulations, defined in terms of a static potential
measured by Wilson loops, is related to continuum physics. 
Hence one does not know e.g. the $\msbar$ gauge coupling $g^2(\mu)$
at $\mu=m_W$ to which some given 4d simulation
corresponds. Third, there are many different length
scales in the problem ---
$(2\pi T)^{-1}, m_W^{-1}(T)$ and $m_H^{-1}(T)$ ---
making the simulations technically demanding. In particular, 
for a large pole Higgs mass $m_H$ the 
inverse correlation length $m_H^{-1}(T)$ at the 
phase transition point becomes much larger than 
$(2 \pi T)^{-1}$, necessitating very large lattices. 

The 3d approach overcomes most of the direct problems of
the 4d approach. Chiral fermions appear only in the perturbative
dimensional reduction step and hence the final effective theory is purely
bosonic. The 3d gauge coupling of the effective theory is in direct
perturbative relation to the original zero temperature 4d gauge
coupling in $\msbar$ or any other desired scheme (in the 
Standard Model, it is related directly to the muon lifetime). 
Finally, dimensional
reduction removes the smallest length scale $(2\pi T)^{-1}$ from the problem, 
making the simulations technically less demanding. Hence one
can easily go to larger Higgs masses. In practice, 4d simulations 
have concentrated on $m_H=34$ GeV, whereas in 3d the continuum
limit has been taken up to $m_H\approx 70$ GeV. 

On the other hand, the 3d approach of course relies 
on perturbation theory in the derivation of the effective
theory. While the perturbation theory used is free of 
IR-problems and is hence expected to be as accurate as 
perturbation theory at zero temperature, it might
nevertheless be useful to check its accuracy non-perturbatively. 
In particular, to obtain the simplest possible effective theory, 
one neglects higher-dimensional radiatively generated 
operators with small coefficients, for instance of the type
\be
O_6=c\frac{\phi^6}{T^2}. \la{phi6}
\ee 
In principle, there might also be non-local
operators generated at high loop orders.
The perturbative estimates made in~\cite{fkrs1} 
indicate that the effects of higher-order
operators should be on the 1\% level
(for other estimates, see~\cite{jkp}).
 
The problems and benefits of the two approaches suggest
that one should take a bosonic theory 
resembling the true original theory, and study the accuracy of 
dimensional reduction at a Higgs mass accessible to both ways
of making simulations. Then one can use the 3d theory to 
include fermions and to go to higher Higgs masses.

To investigate the accuracy of dimensional reduction, we will
in this paper compare the results of 4d and 3d simulations for 
the bosonic SU(2)+Higgs model at the zero temperature pole Higgs mass
of about 34 GeV. Hence the U(1) subgroup and the fermions 
of the Standard Model are neglected. Since the Higgs mass is small, 
the transition is strong and the vacuum expectation value
$v/T_c=\langle\phi\rangle/T_c$ of the Higgs field is large in the broken
phase. Consequently, the comparison should be sensitive
to higher dimensional operators of the type in eq.~\nr{phi6}.
\vali

{\bf 3.} The 4d results relevant for the comparison are
taken from~\cite{des3,des4}. For the zero temperature
parameters, an extrapolation to the continuum limit has 
there been taken. The relation of the Higgs mass to the W 
mass is $m_H/m_W=0.422(11)$, which for 
$m_W=80.22$ GeV gives $m_H=33.9(9)$ GeV (the number
in parentheses is the error of the last shown digit). 
We shall assume $m_H=34(1)$ GeV. The renormalized 
gauge coupling defined in~\cite{des1} has 
the value $g_R^2(m_W^{-1})=0.585(10)$.
In principle, to get the corresponding continuum coupling
one should calculate the static potential from which 
$g_R^2$ is extracted also in the $\msbar$ scheme. 
This kind of a calculation is not available; nevertheless, 
one might expect that $g_R^2(m_W^{-1})=0.585(10)$ roughly
corresponds to the $\msbar$
running coupling $g^2(m_W)\sim 0.585$, since
loop corrections should be of the relative
magnitude $g^2/(16\pi^2)$. 
We shall in the following consider the 
possibilities that $g^2(m_W)=0.570, 0.585, 0.600$.

For the properties
of the phase transition, the extrapolation to the continuum
limit is not quite perfect yet. The critical temperature
has been extrapolated to be $T_c/m_H=2.147(40)$, but the fit
is much affected by the point where the temporal extension
of the lattice is $L_t=2$, relatively far from 
the continuum limit $L_t=\infty$. The latent heat has been measured
to be $L/T_c^4=0.240(34)$ at $L_t=2$ and 
$L/T_c^4=0.28(12)$ at $L_t=4$. For the surface tension, 
values are given only for $L_t=2$; there two different
methods give $\sigma/T_c^3=0.053(5)$ and
$\sigma/T_c^3=0.065(10)$~\cite{des3}. For the order parameter
$v/T_c$ the number 1.65(1) can be read from fig.~9 of~\cite{des4}, 
but there are large systematic errors since the 
relation of the lattice observable
to continuum values is unclear (it should
be noted that $v$ is defined as 
$v^2 \propto \langle \phi^\dagger\phi\rangle$,
leading to a gauge independent but scheme dependent quantity). 
The 4d results used for the comparison are 
summarized in Table~1.
\vali

\begin{table}[t]
\centering

\begin{tabular}{|l|l|l|l|l|l|}
\hline %\cline{3-5}
\mbox{ } $g_R^2(m_W^{-1})$ \mbox{ } & $m_H/$GeV & $T_c$/GeV   
& $L/T_c^4$ & $\sigma/T_c^3$ & $\tilde{v}/T_c$ \\ \hline
\mbox{ } 0.585(10) & 34(1) & 73.0(14) &  0.28(12) &
0.053(5) & 1.65(1) \\
\mbox{ }  & & ($L_t=\infty$) & ($L_t=4$) & 
($L_t=2$) & ($L_t=4$) \\ \hline
\end{tabular}
\vspace*{3mm}

\begin{minipage}[t]{16cm}
\setlength{\baselineskip}{0.6cm}
Table 1: The 4d lattice results for the phase transition~\cite{des3,des4}.
The fit to the continuum limit 
$L_t=\infty$ has been 
studied only for $T_c$. 
To avoid confusion in the comparison with 3d results, 
the order parameter in the broken phase is denoted here by $\tilde{v}$ 
(the definitions are different in 4d and 3d).
\end{minipage}
\end{table}

{\bf 4.} The 3d results relevant for the comparison can 
be read from Table~10 in~\cite{klrs2}. In particular, 
the second block gives the lattice values for the relevant
dimensionless 3d observables, and the third block gives 
the corresponding 4d physical quantities. The results of the third 
block are reproduced in Table~2.

\begin{table}[t]
\centering

\begin{tabular}{|l|l|l|l|l|l|}
\hline %\cline{3-5}
\mbox{ } $g^2(m_W)$ \mbox{ } & $m_H/$GeV & $T_c$/GeV   
& $L/T_c^4$ & $\sigma/T_c^3$ & $v/T_c$ \\ \hline
\mbox{ } 0.420 & 29.1(5) & 76.8(5) &  0.200(7) &
[0.071(3)] & 1.74(3) \\ \hline
\mbox{ } 0.420 & 54.4(5) & 132.6(5) &  0.0294(7) &
0.0017(4) & 0.626(8) \\ \hline
\mbox{ } 0.420 & 64.3(5) & 151.2(5) &  0.0194(12) &
? & 0.529(20) \\ \hline
\end{tabular}
\vspace*{3mm}

\begin{minipage}[t]{16cm}
\setlength{\baselineskip}{0.6cm}
Table 2: The 3d lattice results for the phase transition~\cite{klrs2}.
These results represent the 
continuum limit apart from the surface
tension for $m_H=29.1$ GeV, in which case the 
continuum limit was
not investigated. 
\end{minipage}
\end{table}

Two things should be noted from Table~2, relevant
for the comparison with 4d lattice results.
First, the pole Higgs masses studied
do not include 34~GeV. For the sake of the comparison, 
one should hence make an interpolation to this
Higgs mass. Second, the results shown correspond to the 
$\msbar$ gauge coupling $g^2(m_W)=0.420$ (this value is
close to that in the Standard Model). 
Since 4d simulations were made with a different gauge coupling, 
one also has to study how the results depend on $g^2$.
Fortunately, the scaling with $g^2$  
can easily be studied since the same 3d results 
describe simultaneously a large class of different
4d theories.

We shall first explain the precise way of 
doing the interpolation of $m_H$ and the scaling of $g^2$, 
which was used for deriving the numerical values below.  
Then we also explain an approximate procedure for getting
a rough estimate of the numbers without having to resort to  
the precise form of the analytical formulae~\cite{fkrs1} 
of dimensional reduction.
\vali

{\bf 5.} The precise way of doing the interpolation of $m_H$ 
and change in $g^2$ is based on the dimensionless
quantities $x,y_c,\Delta\ell_3,\ell_3^b,\sigma_3$
of the 3d theory and on their relation 
to 4d physics as explained in Sec.~11 of~\cite{klrs2}.
It should be noted that the values measured for these 
quantities in 3d are independent of the value of $g^2$ 
and of the formulae for dimensional reduction.

First, we need an interpolation of 
$y_c,\Delta\ell_3,\ell_3^b,\sigma_3$ to all $x$. 
Since there are only three lattice points and since they 
are rather far from each other, we use perturbation theory 
to get a better fit. For $y_c$ which is close to zero at 
the critical point, we subtract from the lattice value
the 2-loop perturbative value and then fit a parabola to the 
difference. For the other quantities, we calculate 
the ratio of the lattice and perturbative values and
fit a parabola to this. With this procedure, one should 
get a reasonable interpolation for the critical curve $y=y_c(x)$
and for the values of $\Delta\ell_3,\ell_3^b$ along it. 

For the surface tension, there is only 
data point ($m_H=54.4$ GeV; 
the tree-level Higgs mass is then $m_H^*=60$ GeV) in the 
continuum limit in Table~10 of~\cite{klrs2}, 
and this is pretty far from 
the point $m_H=34$ GeV. Hence, to get some kind of a comparison
also for the surface tension, we include the value of
$\sigma_3$ at $m_H=29.1$ GeV and an estimate
of the surface tension at $m_H^*=70$ GeV from~\cite{leip2} 
into the fit, although these values 
are not extrapolations to the continuum limit. 
The point $m_H^*=70$ GeV does not affect the result
practically at all.

Next, one needs the relations to 4d. We fix $m_H=34$ GeV
and $g^2(m_W)=0.570$, 0.585, 0.600.
For each $g^2$, we vary $T$ and draw the corresponding curve
$x(m_H,T), y(m_H,T)$ in the $(x,y)$-plane. The value of $T$
for which this curve crosses the critical curve $y=y_c(x)$ 
gives the critical temperature. Using
the 3d gauge coupling $g_3^2(m_H,T_c)$, 
the derivatives $dx/dT, dy/dT$ at $T_c$, the slope of the 
critical curve $dy_c/dx$, and the values of 
$\Delta\ell_3,\ell_3^b$ and $\sigma_3$ at $x_c$, one can then 
calculate $v/T_c, \sigma/T_c^3$ and $L/T_c^4$ from 
eqs.~(11.5), (11.6) and (11.7) of~\cite{klrs2}. 

With the method explained, we get the results in Table 3.
We have not added any errors from the interpolation to the values of
$T_c, L/T_c^4$ or $v/T_c$. The error of $T_c$ is mainly determined
by the uncertainty in the Higgs mass, and might be slightly 
underestimated. For $\sigma/T_c^3$ we give just the percentual
error of the $m_H=54.4$ GeV continuum limit; 
if there were simulations at the pole mass 
$m_H=34$ GeV (in which case the tree-level mass parameter
is $m_H^*\approx 41$ GeV) the errors would be considerably smaller.
\vali

\begin{table}[t]
\centering

\begin{tabular}{|l|l|l|l|l|l|}
\hline %\cline{3-5}
\mbox{ } $g^2(m_W)$ \mbox{ } & $m_H/$GeV & $T_c$/GeV  
& $L/T_c^4$ & $\sigma/T_c^3$ & $v/T_c$ \\ \hline
\mbox{ } 0.570 & 34.0(5) & 76.7(5) & 0.19(1)   
& 0.048(12) & 1.43(4)      \\ \hline
\mbox{ } 0.585 & 34.0(5) & 75.7(5) & 0.19(1)   
& 0.050(12) & 1.44(4)      \\ \hline
\mbox{ } 0.600 & 34.0(5) & 74.9(5) & 0.20(1)   
& 0.052(13) & 1.45(4)      \\ \hline
\end{tabular}
\vspace*{3mm}

\begin{minipage}[t]{16cm}
\setlength{\baselineskip}{0.6cm}
Table 3: The 3d lattice results interpolated to $m_H=34$ GeV at 
different gauge couplings. These results represent the 
continuum limit apart from the surface tension (see Table~2).
\end{minipage}
\end{table}

%\noindent

{\bf 6.} For a rough estimate of the numbers of Table 3, 
one can use a simplified procedure. 
First, just interpolate the values in Table 2
to $m_H=34$ GeV with the method explained above, giving $T_c=88.4(5)$ GeV, 
$L/T_c^4=0.118(5), \sigma/T_c^3=0.031(8), v/T_c=1.30(3)$. 
Then look from Sec.~11 of~\cite{klrs2} how the values
scale with $\hat{g}_3^2\equiv 
g_3^2/T_c\approx g^2(7T_c)$. The critical temperature
is determined approximately from
\be
m_3^2\sim -\frac{\bar{m}_H^2}{2} + c \hat{g}_3^2 T_c^2=0, \la{Tc}
\ee 
where $m_3^2$ is the scalar mass parameter in the dimensionally
reduced 3d theory, $c$ is a constant, and $\bar{m}_H^2$ is the 
running $\msbar$ Higgs mass at a scale of order $7T_c$.
It follows from eq.~\nr{Tc} 
that $T_c\propto \hat{g}_3^{-1}$. From eqs.~(11.5) and (11.6) 
one can see that
$v/T_c$ roughly scales as $\hat{g}_3$ and
$\sigma/T_c^3$ scales as $\hat{g}_3^4$. 
To get the scaling of $L/T_c^4$, one can neglect the latter 
term in eq.~(11.7) since $dx/dT$ is non-zero only through
logarithmic one-loop corrections to the dimensional
reduction of the coupling constants. Using 
$y=m_3^2/g_3^4$ one then gets
\be
\frac{dy}{dT}\sim\frac{d}{dT}\biggl(
\frac{-\bar{m}_H^2/2 + c \hat{g}_3^2 T^2}
{\hat{g}_3^4T^2}\biggr)\approx \frac{\bar{m}_H^2}{g_3^4 T_c}
\approx \frac{2 c}{g_3^2},
\ee
where the logarithmic running of $\hat{g}_3^2$ and $\bar{m}_H^2$ 
with temperature was neglected and eq.~\nr{Tc} was used. 
Using eq.~(11.7), one finally gets $L/T_c^4\propto \hat{g}_3^4$.

The values of $\hat{g}_3^2$ are $\hat{g}_3^2=0.384,0.521$ 
for $g^2(m_W)=0.420, 0.585$, respectively. This gives
$T_c=75.9(5)$ GeV, 
$L/T_c^4=0.22(1), \sigma/T_c^3=0.057(15), v/T_c=1.51(4)$
for $g^2(m_W)=0.585$.
These values are in rough agreement with  
the second row of Table~3.
\vali

{\bf 7.} Let us finally compare the 4d results in Table~1 
and the 3d results in Table~3. For the critical temperature,
the 3d values are a bit higher, but compatible within 
errorbars for $g^2(m_W)=0.600$. This might be an indication
of the correct value of $g^2(m_W)$. A higher value of
$g^2$ also brings the other 
variables closer to each other.
One should also remember that there may be small errors e.g.\
in the interpolation. In addition, it should be reiterated that 
the extrapolation of the 4d critical temperature to the continuum 
limit might allow for improvement.

The values of latent heat are clearly compatible, but the 
errorbars in 4d are rather large. The values
of surface tension agree, as well, but the errorbars
are again large, and neither the 4d nor the 3d values 
represent the continuum limit at this $m_H$.
In $v/T_c$ there is some
discrepancy which is not surprising since the definitions
of $v/T$ are different in 4d and 3d.
 
We conclude that at the present level of accuracy, 
the results agree. This gives non-perturbative
evidence for the 
applicability of dimensional reduction in the present context. 
%Moreover, the 3d approach produces smaller errorbars in the
%continuum limit at least for $T_c$ and $L/T_c^4$.

Clearly, it would be most desirable to improve the 
quality of the comparison. Especially the 
extrapolation to the continuum limit should be made more
precise. In the 4d case $T_c$ relies perhaps too much
on $L_t=2$, and for $L/T_c^4$ and $\sigma/T_c^3$
a fit to $L_t=\infty$ is lacking altogether. In the 
3d case the fit to the 
continuum limit for $\sigma/T_c^3$ at 
$m_H=54.4$~GeV ($m_H^*=60$~GeV)
contains large errors since the transition is already 
rather weak there; for $m_H=29.1$~GeV 
($m_H^*=35$~GeV) the continuum 
limit is much easier but was not studied in~\cite{klrs2}. 
Of course, it would
also be most important to get some direct evidence
on the value of $g^2(m_W)$ to which the 4d simulations
correspond. A comparison of finite temperature
correlation lengths, which seems not to be available
at the moment, would be welcome.
Finally, all the comparisons
would be more straightforward and the errors
related to interpolation removed if 
lattice simulations were made with the same 
pole Higgs masses. 
%Interpolation is especially questionable
%for the surface tension which deviates much from the 
%perturbative value.
Once the errors have been reduced, 
%it will very interesting to redo the comparison
%to reliably estimate 
the effects of higher-order operators can be reliably estimated. 
%dimensional reduction.
\vali

{\bf Acknowledgements.} 
This investigation was prompted and improved by
discussions with K.Kajantie, A.Laser, M.G.Schmidt and M.Shaposhnikov. 
I also thank Z.Fodor and K.Rummukainen for useful discussions. 
The investigation was partially supported by the University of Helsinki.

\end{document}